\documentclass{emulateapj}

\usepackage{epsfig}

\newcommand{\be}{\begin{equation}}
\newcommand{\ee}{\end{equation}}
\newcommand{\ba}{\begin{eqnarray}}
\newcommand{\ea}{\end{eqnarray}}

\begin{document}

\pagestyle{plain}

\title{Effects of photometric redshift uncertainties on weak lensing tomography}

\author{Zhaoming Ma{\footnote{Email: mzm@oddjob.uchicago.edu}}, Wayne Hu and
        Dragan Huterer}
\affil{Kavli Institute for Cosmological Physics 
       and Department of Astronomy \& Astrophysics,\\
        University of Chicago, Chicago, IL 60637}

\begin{abstract}

We perform a systematic analysis of the effects of photometric redshift
uncertainties on weak lensing tomography. We describe the photo-$z$
distribution with a bias and Gaussian scatter that are allowed to vary 
arbitrarily between intervals of $\delta z = 0.1$ in redshift.  
While the mere presence of bias and scatter does not substantially degrade
dark energy information, uncertainties in both parameters do.
For a fiducial next-generation survey each would need to be known to 
better than about $0.003-0.01$ in redshift for each 
interval in order to lead to less than a factor of $1.5$ increase in the dark 
energy parameter errors. The more stringent requirement corresponds to a 
larger dark energy parameter space, when redshift variation in the equation 
of state of dark energy is allowed.  Of order $10^4-10^5$ galaxies with 
spectroscopic redshifts fairly sampled from the source galaxy distribution
will be needed to achieve this level of calibration. If the sample is
composed of multiple galaxy types, a fair sample would be required for each.
These requirements increase in stringency for more ambitious surveys; we 
quantify such scalings with a convenient fitting formula. No single aspect 
of a photometrically binned selection of galaxies such as their mean or 
median suffices, indicating that dark energy parameter determinations are 
sensitive to the shape and nature of outliers in  the photo-$z$ redshift 
distribution.
\end{abstract}

\section{Introduction}
\label{sec:introduction}

Weak gravitational lensing of galaxies by large scale structure is rapidly
becoming one of the most powerful cosmological probes
\citep{Bart_Schneider,Refregier}.  Following the first detections a few years
ago \citep{vW_first,Kaiser_first,Bacon_first,Wittman_first,RRG}, weak lensing
has produced increasingly better constraints on the matter density relative to
critical $\Omega_m$ and the amplitude of mass fluctuations $\sigma_8$
\citep{Hoekstra,Pen_vW_Mel,Brown,Jarvis_etal_2003,Pen_etal,Heymans,vW_VIRMOS}.
While weak lensing is most sensitive to the amount and distribution of dark 
matter, it also has the potential to probe the dark energy through its effect 
on the growth of structure and distances \citep{Hu_Tegmark,Huterer02,Hu_tomo2,
Takada_Jain,Song_Knox,Ishak}.  Indeed, when combined with other cosmological 
probes, weak lensing data already produce interesting constraints on the dark 
energy \citep{Jarvis_etal_2005}. 

By utilizing source galaxy redshifts to study the growth of structure and the
distance-redshift relation tomographically, substantially more dark energy 
information can be recovered \citep{Hu_tomo}. In fact future weak lensing 
surveys such as PanSTARRS\footnote{http://pan-starrs.ifa.hawaii.edu}, 
Supernova/Acceleration Probe \citep[SNAP\footnote{http://snap.lbl.gov};][]
{SNAP} and Large Synoptic Survey Telescope (LSST\footnote{http://www.lsst.org})
are expected to impose constraints on dark energy that are comparable to 
those from type Ia supernovae \citep[see e.g.\ ][]{wl_space_III}.  In the 
more near term, the Canada-France-Hawaii Telescope Legacy Survey 
(CFHTLS\footnote{http://www.cfht.hawaii.edu/Science/CFHLS}) and the Dark Energy
Survey\footnote{http://cosmology.astro.uiuc.edu/DES} are expected to help bridge
the gap between the current and ambitious future surveys.

Powerful future surveys will require a much more stringent control of the
systematics.  Recent work has addressed systematic errors from the
computation of  the non-linear power spectrum \citep{Vale_White,
White_Vale, LosAlamos, Huterer_Takada, Hagan_Ma_Kravtsov}, baryonic cooling 
and pressure forces on the distribution of large-scale structures 
\citep{White_baryons, Zhan_Knox}, approximations in inferring the shear from 
the maps \citep{Dod_Zhang, White_reduced}, the presence of dust 
\citep{Vale_Hoekstra}. Such studies have stimulated work on how to improve 
the PSF reconstruction \citep{Jarvis_Jain}, estimate shear from noisy maps 
\citep{Bernstein_Jarvis,Hirata_Seljak,Hoekstra04}, and protect against the
small-scale biases in the power spectrum \citep{nulling}.

In this work we consider the effect of errors in photometric redshifts of
source galaxies on weak lensing tomography.  Of course,
the total number of galaxies, which is currently in the millions and might be
in the billions with future surveys, is too large for spectroscopic measurements
to be feasible. One therefore needs to rely on the photometric redshifts whose
accuracy with the current state of the art in photometry, algorithms, galaxy
classification etc, while presently adequate, may not be sufficient for future 
surveys which are expected to have very small statistical errors.  Uncertain
photometric redshifts blur the tomographic bin divisions of source galaxies. In
the extreme case when photometric redshift errors are comparable to the
width of the distribution itself, one completely loses tomographic information 
degrading the cosmological parameter accuracies by up to an order of magnitude.

In this paper we study how the photometric redshift uncertainties affect
cosmological parameter determinations.  We construct an explicit mapping between
the photometric and true redshifts, and parametrize it to allow an arbitrary
evolution of the bias and scatter  between discrete redshift intervals.  
We then study
how accurately the photometric redshifts need to be known {\it a priori} and,
in particular, which details of the photometric redshift error distribution are
the main source of degeneracy with cosmological parameters.   We hope that this
study will help stimulate work on assessing and improving existing algorithms 
for photometric redshift estimation \citep[e.g][]{cunha}.

The outline of the paper is as follows. In {\S}\ref{sec:methodology}, we 
introduce the formalism and parametrizations of both cosmology and 
photometric redshift errors.   We explore the loss of lensing information 
on the dark energy to photometric redshift uncertainties in 
\S \ref{sec:information}.  We show how this lost information is regained as 
we impose prior knowledge of the photometric redshift parameters in 
\S \ref{sec:results}.  We discuss our results and conclude in 
{\S}\ref{sec:discussion}.

\section{Methodology}
\label{sec:methodology}

In this section, we discuss the modeling of the photometric redshift
distribution.  We then illustrate the flexibility of this description
through two different fiducial models for the
distribution. Finally we discuss its relationship to lensing observables and
the Fisher formalism for addressing its impact on parameter estimation.

\subsection{Photo-$z$ Distribution}
\label{sec:photozdistribution}

  Having only the photometric redshift (``photo-$z$") of the source galaxies 
at hand, the observer will necessarily bin the galaxies by their photometric 
redshifts $z_{\rm{ph}}$ rather than true (spectroscopic) redshifts $z$.  
With a probability distribution $p(z_{\rm{ph}} | z)$ in $z_{\rm ph}$ at
a given $z$, the true redshift distributions of the bins necessarily overlap.  

 In general this distribution can vary arbitrarily with $z$.  The true  
distribution of galaxies $n_i(z)$ that fall in the $i$th photo-$z$ bin with 
$z_{\rm{ph}}^{(i)}<z_{\rm{ph}}<z_{\rm{ph}}^{(i+1)}$ becomes
\be n_i(z) = \int_{z_{\rm{ph}}^{(i)}}^{z_{\rm{ph}}^{(i+1)}}
dz_{\rm{ph}} \, n(z)\, p(z_{\rm{ph}} | z)\,.
  \label{eq:ni}
\ee
$n(z)=d^2N /dz d\Omega$ is the overall galaxy redshift distribution, and is
chosen to have the form
\be
n(z) \varpropto {z}^{\alpha} \exp\left [-(z/z_0)^\beta\right ] \,.
\label{eq:nz}
\ee
Unless otherwise stated we will adopt $\alpha=2$, $\beta=2$ and fix $z_0$ such
that median redshift is $z_{\rm med} = 1$.  The  total number of galaxies per 
steradian
\be
n^A = \int_0^\infty dz n(z) \,,
\ee
fixes the normalization, and we analogously define
\be
n^A_{i} = \int_0^\infty dz n_{i}(z) \,
\ee
for the bins.

By construction, the sum of the individual distributions equals the total 
$\sum_i n_i(z) = n(z)$. Therefore, regardless of how complicated the photo-$z$ 
probability distribution gets and hence the redshift distributions of the 
tomographic bins, the total distribution of galaxies $n(z)$ is unchanged.  

This construction cleanly separates uncertainties due to the photometric
redshifts of the individual survey galaxies characterized by 
$p(z_{\rm{ph}} | z)$ from uncertainties in the redshift distribution of the
underlying total distribution of galaxies $n(z)$.  We mainly consider  the 
former in this work but comment on the latter in \S \ref{sec:discussion} 
\citep[see also][]{wlsys, Ishak_Hirata}.  The rationale is that even without 
any knowledge of the photo-$z$'s of the
survey galaxies themselves, one can at least bin all of the galaxies together
assuming that the underlying redshift distribution or selection function of the
survey is known.  In practice, this means that one must obtain information
about the underlying distribution from an independent source (say, another
survey through a study of the luminosity function) or from a fair subsample of
survey galaxies with spectroscopic redshifts.

\begin{figure}[ht]
\centerline{\psfig{file=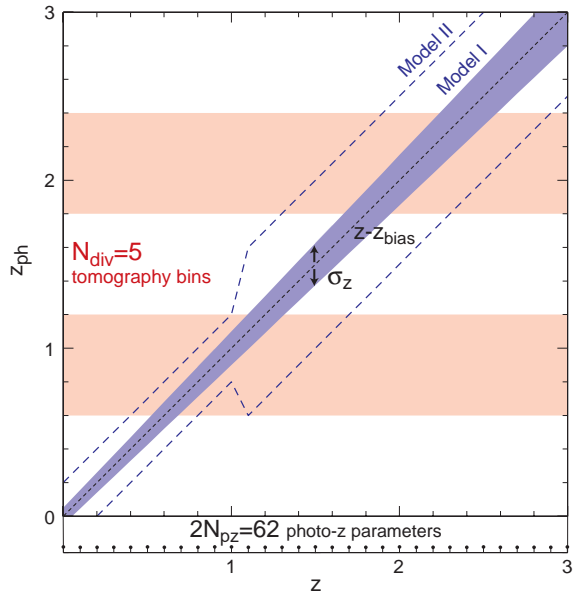, width=3.0in}}
  \caption{Parametrization of the photo-$z$ distribution 
    and two illustrative fiducial models.  The distribution spreads galaxies
    at a given redshift $z$ into a distribution in $z_{\rm ph}$ characterized 
    here by a bias $z_{\rm bias}$ and a scatter $\sigma_z$ whose evolution is 
    parametrized by interpolating their values at $N_{\rm pz}$ redshifts 
    $z_{\mu}$.  In both models $z_{\rm bias}=0$, whereas $\sigma_z$, given in 
    \S \ref{sec:photozmodel}, is illustrated here for model I (shaded region) 
    and model II (dashed lines) as 1$\sigma$ bands.  Galaxies binned according 
    to their photometric redshifts ($N_{\rm div}$ horizontal bands) then have 
    overlapping redshift distributions determined by the $2 N_{\rm pz}$ 
    photo-$z$ parameters.}
  \label{fig:mapping}
\end{figure}

\subsection{Photo-$z$ Models}
\label{sec:photozmodel} 

Any photo-$z$ model may be described by providing a function for the
distribution of photometric redshifts given the true redshift, $p(z_{\rm{ph}} |
z)$.  For the purposes of this paper we take the simplifying assumption that
this function is a Gaussian at each redshift, i.e.\
\be
p(z_{\rm ph} | z) = 
  {1 \over {\sqrt{2\pi} \sigma_z}}
  \exp\left [-{{(z-z_{\rm{ph}} -z_{\rm bias})^2} \over {2 {\sigma_z}^2}}\right ].
  \label{eq:f}
\ee 
However, we allow the bias $z_{\rm{bias}}(z)$ and scatter $\sigma_{{z}}(z)$ to 
be arbitrary functions of redshift.  The redshift distribution of the 
tomographic bins defined by equation~(\ref{eq:ni}) can then be written as
\begin{eqnarray}
 n_i(z) &=&{} \onehalf n(z)\, [ {\rm erf}(x_{i+1}) - {\rm erf}(x_i) ], 
 \end{eqnarray}
 with
\begin{eqnarray}
   x_i &\equiv & ({{z_{\rm{ph}}^{(i)}-z+z_{\rm{bias}}})/{\sqrt{2} \sigma_z}},
\end{eqnarray}
where ${\rm erf}(x)$ is the error function.

The Gaussian assumption is not as restrictive as it might naively seem.  By
allowing the bias and scatter to be arbitrary functions of redshift one can
obtain arbitrarily complex redshift distributions in the tomographic bins
through equation~(\ref{eq:ni}).  In fact, the mapping is in principle completely
general for finite bins and a smooth underlying distribution.  Galaxies in a
finite range of redshift over which the distribution is nearly constant can
then be mapped to any $z_{\rm{ph}}$.

In practice we will represent the free functions $z_{\rm{bias}}(z)$ and
$\sigma_{z}(z)$ with a discrete set of $N_{\rm pz}$ photo-$z$ parameters. 
They represent the values of the functions
 at $z_{\mu}$ which are equally spaced from $z=0$ to 3.
 To evaluate the functions at an arbitrary redshift, we
take a linear interpolation of the discrete parameters in redshift.

\begin{figure}[ht]
\centerline{\psfig{file=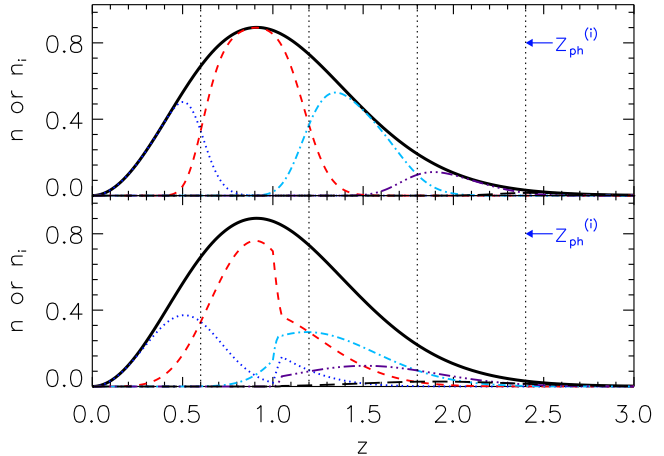, width=3.5in}}
  \caption{Source galaxy redshift distribution $n(z)$. Top panel: photo-$z$
           model I. Lower panel: photo-$z$ model II. The solid 
           curve is the overall galaxy distribution
           defined in equation~(\ref{eq:nz}).  The other
           curves are the true (spectroscopic) distributions that
           correspond to the sharp divisions in photo-$z$ space
           (denoted by dotted vertical lines). }
% Ploted using cheops4 0112 and 0116 by nzFinal.pro
  \label{fig:nz}
\end{figure}

While a finite $N_{\rm pz}$ does restrict the form of the distribution, it
still allows radically different redshift distributions given the same
tomographic bins.  For example, consider two different photo-$z$ models
\begin{itemize}
\item Model I: $z_{\rm bias}(z)=0$; $\sigma_z(z) = 0.05(1+z)$.  
\item Model II: $z_{\rm bias}(z)=0$;
 $\sigma_z(z) = 0.2$ for $z < 1.0$ and 
  $\sigma_z(z) = 0.5$ for $z > 1.0$. 
\end{itemize}
The distribution $p(z_{\rm ph}|z)$ is illustrated in Figure~\ref{fig:mapping} 
for $N_{\rm pz}=31$ through the $1\sigma$ scatter region.  The resulting 
redshift distributions for $N_{\rm div}=5$ tomographic bins are shown in
Figure~\ref{fig:nz}.  These specific choices of $N_{\rm pz}$ and $N_{\rm div}$ 
are motivated in \S \ref{sec:information}.

  Model II demonstrates that sharp changes in the Gaussian
photometric parameters can map neighboring galaxies in redshift to quite
different tomographic bins.  The redshift distributions of the bins
can thus have features that are sharper than the assumed scatter.
Additionally, photo-$z$ degeneracies that take two distinct spectroscopic
redshift ranges into a single photometric redshift and lead to bimodality in 
the binned distribution can be modeled by a large $z_{\rm bias}$.
Finally, galaxy types with different photo-$z$ distribution at a given redshift
can be approximated by discontinuous jumps between infinitesimally spaced 
redshift bins. Such considerations, while potentially important are beyond the 
scope of this work. 

In summary allowing our set of parameters to freely
vary one can access a wide range of tomographic redshift distributions.
Uncertainty in these parameters will then cause uncertainties in tomographic
dark energy determinations.

\subsection{Lensing Observables}
\label{sec:observables}

The convergence power spectrum at a fixed multipole $\ell$ and for the $i$th 
and $j$th tomographic bin $P_{ij}^{\kappa}(\ell)$
 is given by \citep{Kaiser_92, Kaiser_98}
\begin{equation}
n^A_i n^A_j P_{ij}^{\kappa}(\ell) = 
\int_0^{\infty} dz \,{W_i(z)\,W_j(z)}{ H(z) \over D^2(z)}\,
 P\! \left (k_{\ell},  z\right ),
\label{eq:pk_l}
\end{equation} 
\noindent where $H(z)$ is the Hubble parameter, and $D(z)$ is the angular
diameter distance in comoving coordinates. $P(k_{\ell}, z)$ is the
three-dimensional matter power spectrum and $k_{\ell} = \ell /D(z)$ is the
wavenumber that projects onto the multipole $\ell$ at redshift $z$.  The
weights $W$ are given by
\begin{eqnarray}
 W_i(z) &=& {3\over 2}\,\Omega_m\, {H_0^2 D(z) \over H(z)}(1+z)\nonumber  \\
 &&\times 
 \int_z^\infty 
 dz^{\prime} {n_i(z^{\prime})}
{D_{LS}(z,z') \over D(z')} \,,
\label{eqn:weights}
 \end{eqnarray}
 where $D_{LS}(z,z')$ is the angular diameter distance between the two 
redshifts.  The power spectrum is computed from the transfer function of 
\cite{Hu_transfer} with dark energy modifications from \cite{Hu01c}, and the 
non-linear fitting function of \cite{PD96}.
 
With tomographic binning, the number weighted power spectrum 
$n^A_i n^A_j P_{ij}^{\kappa}$ and not $P_{ij}^{\kappa}$ is the fundamental 
observable.  Even given photometric redshift uncertainties in the binning it 
is always possible to recover the total weighted power spectrum \citep{Hu_tomo}
\begin{equation}
(n^{A})^{2}P^{\kappa} = 
\sum_{i,j = 1}^{N_{\rm div}} n^A_i n^A_j P^{\kappa}_{ij} \,,
\label{eq:ninjpij}
\end{equation}
since the weighting is based on the observed $n^{A}_{i}$.  By treating 
$n^A_i n^A_j P_{ij}^{\kappa}$ as the observable one guarantees that the
addition of photo-$z$ estimates for the individual galaxies can only
add information.    This would not be true if $P_{ij}^{\kappa}$ were taken as
the only observable quantity.  Given that  changes in photo-$z$ parameters
induce changes in $n^{A}_{i}$, the binned power spectra $P_{ij}^{\kappa}$ do not
contain enough information to weight the power spectra and 
recover the total $P^{\kappa}$.

That the binned angular number densities
 $n^{A}_{i}$ are observed quantities also implies that there is additional
direct information on the photo-$z$ parameters that does not
depend on shear measurements.  For example, a high fraction of
galaxies in bins with $z_{\rm ph}$ larger than the median redshift would imply
a large photo-$z$ bias.  We choose not to consider this sort of information 
since it is not directly related to lensing.  Furthermore for the small
changes in $n^{A}_{i}$ that we will typically be considering, the sample
variance between the observed $n^{A}_{i}$ and that predicted by the
underlying redshift distribution and the photo-$z$ parameters cannot
be ignored \citep{Hu_Kravtsov}.   Therefore we will consider the
number weighted power spectra 
$n^{A}_{i} n^{A}_{j} P_{ij}^{\kappa}$ as the fundamental lensing observables.

\subsection{Fisher Matrix}
\label{sec:fisher}

The Fisher matrix quantifies the information contained in the lensing 
observables
\be
O_{a=i(i-1)/2+j}(\ell) = n^A_i n^A_j P_{ij}^{\kappa}(\ell)\,, \quad (i\ge j)
\ee
on a set of cosmological and photo-$z$ parameters $p_{\mu}$.  Under the 
approximation that the shear fields are Gaussian out to $\ell_{\rm max}$, 
the Fisher matrix is given by
\be
F_{\mu \nu} = \sum_{\ell=2}^{\ell_{\rm max}}(2\ell+1)f_{\rm sky}{\sum_{ab}
             {\partial{O_{a}}
            \over \partial{p_{\mu}} } [ {\bf C}^{-1} ]_{ab} 
             {\partial{O_{b}}
            \over \partial{p_{\nu}}  }}\,,
\label{eq:fisher}
\ee
so that the errors on the parameters are given by $\Delta p_{\mu} = [{\bf F}^{-1}]^{1/2}_{\mu\mu}$.

Given shot and Gaussian sample variance, the covariance matrix  of the
observables becomes 
\be
C_{ab} = 
n^{A}_{i} n^{A}_{j} n^{A}_{k} n^{A}_{l} \left(
P^{\rm tot}_{ik}  P^{\rm tot}_{jl} + 
        P^{\rm tot}_{il}  P^{\rm tot}_{jk}\right)\,,
\label{eq:Cov}
\ee
where $a = i(i-1)/2+j$, $b= k(k-1)/2+l$.  The total power spectrum is given by 
\begin{equation}
P^{\rm tot}_{ij}=P_{ij}^{\kappa} + 
\delta_{ij} {  \gamma_{\rm int}^2  \over {n}_i^A} \,,
\label{eq:C_obs}
\end{equation}
where $\gamma_{\rm int}$ is the rms shear error per galaxy per component 
contributed by intrinsic ellipticity and measurement error.
For illustrative purposes we will use $\ell_{\rm max}=3000$,
$f_{\rm sky}$ corresponding to 4000 sq.\ deg, $\bar n^{A}$ corresponding to 
55 galaxies/arcmin$^{2}$ and $\gamma_{\rm int}=0.4$.
The value of $\ell_{\rm max}$ is motivated by simulations which find
substantial deviations from Gaussianity and weak lensing approximations
on arcminute scales \citep{White_Hu,Vale_White}.  For our choice of  noise
parameters above, the results are insensitive to $\ell_{\rm max}$ since
the measurements are noise dominated on those scales.

For the cosmological parameters, we consider four parameters that affect the
matter power spectrum: the physical matter density $\Omega_{m} h^{2} (= 0.14)$, 
physical baryon density
$\Omega_{b} h^{2} (= 0.024)$, tilt $n_{s} (= 1)$, and the amplitude 
$\delta_{\zeta}(=5.07 \times 10^{-5}$ ; or $A= 0.933$ \citet{Speetal03}).
Values in parentheses are those of 
the fiducial model. Unless otherwise stated, we shall take priors on these 
four parameters of $\Delta \ln \Omega_m h^2=\Delta \ln \Omega_b h^2 =
\Delta \ln \delta_\zeta= \Delta  n_s=0.05$.  These priors represent only 
a modest improvement over current determinations. Our results on the relative 
degradation in constraints caused by photo-$z$ errors are 
insensitive to reasonable variations in this choice.

To these four cosmological parameters, we add either two or three dark 
energy parameters: the dark energy density $\Omega_{\rm DE}(=0.73)$, 
its equation of state today $w_{0} = p_{\rm DE}/\rho_{\rm DE}|_{z=0} (=-1)$ 
and optionally its derivative $w_{a} = -dw/da |_{z=0} (=0)$ 
assuming a linear evolution with the scale factor $w = w_{0} + (1-a)w_{a}$.

Note that throughout this paper, our notational convention is
latin indices for tomographic bins and greek indices for parameters.

\begin{figure}[ht]
  \centerline{\psfig{file=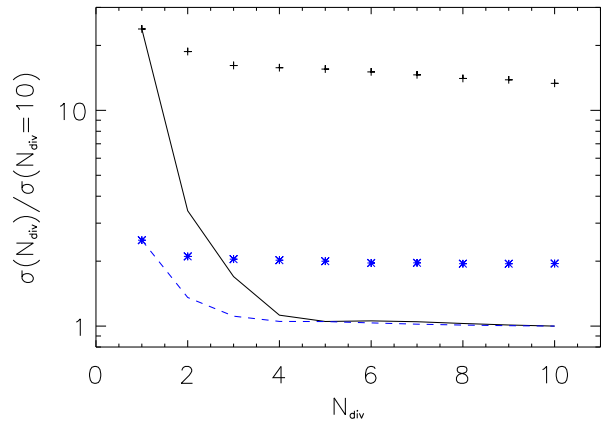, width=3.2in}}
  \caption{Relative errors in dark energy parameters as a function of the 
           number of tomographic divisions $N_{\rm div}$.  Solid lines 
           correspond to $w_a$ for the set $\{w_0,w_a,\Omega_{\rm{DE}}\}$; 
           dashed lines to $w_0$ for  $\{w_0,\Omega_{\rm{DE}}\}$.
           Both lines assume that all photo-$z$ parameters are perfectly 
           known (i.e.\ fixed).  Note that the results converge at smaller
           $N_{\rm{div}}$ for a smaller dark energy space, and that
           $N_{\rm{div}}=5$ is more than sufficient in either case.  The points
           correspond to the same cases, but now with $2 N_{\rm pz}=62$ 
           photo-$z$ parameters marginalized (Photo-$z$ priors of
           $\Delta \sigma_z = \Delta z_{bias} = 10 $ are applied).
           Here essentially all tomographic information is lost so 
           that the errors are comparable to those of $N_{\rm div}=1$ or no 
           tomography.}
  % Plotted using numDivNew1.pro (cheops4)
  \label{fig:numDiv}
\end{figure}

\section{Dark Energy Information Loss}
\label{sec:information}

In this section we consider the nature of the tomographic information on the 
dark energy and its loss to photo-$z$ uncertainties.  We establish the maximal 
information that can be gained through tomographic redshift divisions
for a given dark energy parametrization.  We then determine the number
of photo-$z$ degrees of freedom that would be required to lose this
information.   This loss of information is caused by a degeneracy between 
cosmological and photo-$z$ parameters.   We explicitly construct an example 
of this degeneracy as both an illustration and test of our statistical 
methodology.

\subsection{Maximal Information and $N_{\rm div}$}

For any given choice of dark energy parameterization, the information contained
in lensing will saturate with some finite number of tomographic bins 
$N_{\rm div}$ \citep{Hu_tomo}. Since the broad lensing kernel of 
equation~(\ref{eqn:weights}) makes the shear for neighboring source redshifts 
highly correlated, most of the information is contained in a few
coarse bins.  The exact number depends on the type of information that is to
be extracted.  Roughly speaking, the number of bins should exceed the number
of dark energy parameters.  

Figure~\ref{fig:numDiv} (lines) quantifies this expectation through improvement 
in the errors on dark energy parameters as a function of $N_{\rm div}$ for 
Model I.  For a 2 parameter dark energy space 
 $\{w_0,\Omega_{\rm{DE}}\}$,  $N_{\rm{div}}=3$ divisions equally spaced from 
$z=0$ to $z=3$ are enough for the improvements in $w_0$ to saturate.  For a 3 
parameter dark energy space  $\{w_0,w_a,\Omega_{\rm{DE}}\}$,  $N_{\rm{div}}=4$ 
divisions are sufficient for $w_a$. Note that $N_{\rm div}=1$ corresponds to 
no tomography or no photo-$z$ information on the individual galaxies.  
The dark energy parameters that are not shown in Figure~\ref{fig:numDiv} 
behave similarly.  In what follows we conservatively
adopt $N_{\rm{div}}=5$ as sufficient to extract the dark energy information.
With $N_{\rm{div}}=5$ and photo-$z$ parameters fixed, the constraints on
dark energy parameters are shown in Table\,\ref{tbl-1}.

\begin{deluxetable}{ccllcc}
\tablewidth{0pt}
\tablecaption{Baseline constraints on dark energy parameters \label{tbl-1}}
\tablehead{
\colhead{Photo-$z$ Model}              & \colhead{Parameters}             &
\colhead{$\sigma{(\Omega_{\rm DE})}$}  & \colhead{$\sigma{(w_0)}$}        &
\colhead{$\sigma{(w_a)}$} }
\startdata
I  & $\{\Omega_{\rm DE}, w_0\}$      & 0.0062 & 0.061 & -    & \\
   & $\{\Omega_{\rm DE}, w_0, w_a\}$ & 0.024  & 0.25  & 0.69 & \\
II & $\{\Omega_{\rm DE}, w_0\}$      & 0.0073 & 0.070 & -    & \\
   & $\{\Omega_{\rm DE}, w_0, w_a\}$ & 0.034  & 0.36  & 0.96 & 
\enddata
\end{deluxetable}

Note that improvements relative to the no-tomography case are more significant 
in the larger parameter space.   This is due to the fact that $w_{0}$ is
nearly degenerate with $w_{a}$ since lensing mainly constrains $w(z)$ at some
intermediate redshift (see below).  Even the small amount of information in the
fine-binned tomography assists the breaking of the degeneracy.

\subsection{Maximal Degradation and $N_{\rm{pz}}$}
\label{sec:Npz}

Next we choose the number of photo-$z$ parameters $N_{\rm pz}$ that describe 
each of the functions  $z_{\rm bias}(z)$ and $\sigma_{z}(z)$.  We seek to 
allow enough freedom in the photo-$z$ parameters so that in the absence of 
prior information on their values all of the tomographic information is lost.   
Because the limit of no tomographic information corresponds to $N_{\rm div}=1$, 
%Because the case of no-tomography or $N_{\rm div}=1$ sets this criteria
we have a quantitative means of assessing the minimal $N_{\rm pz}$.
When $N_{\rm pz}$ becomes large enough, the variations in
redshift, which act on the characteristic scale of 
$\delta z=3.0/(N_{\rm pz}-1)$, are rapid enough that they do not mimic any
variation in cosmological parameters.

\begin{figure}[t]
\centerline{\psfig{file=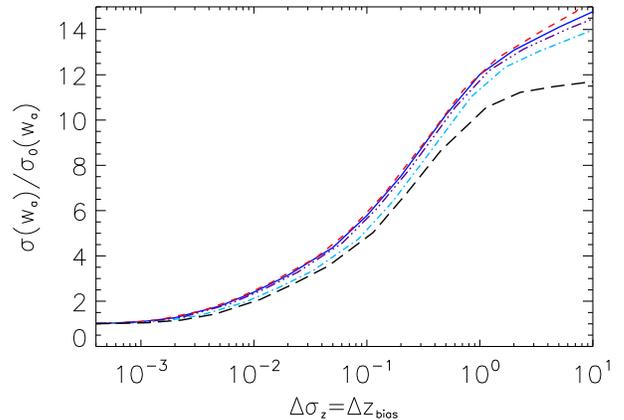, width=3.2in}}
  \caption{Error degradations in $w_a$ (that is, errors in $w_a$ relative to
           the error with perfect knowledge of photo-$z$ parameters) as a
           function of the photo-$z$ prior.  The photo-$z$ priors are rescaled 
           by a factor of $\sqrt{N_{\rm{pz}}/31}$ so that they reflect 
           constraints per $\delta z=0.1$ independently of $N_{\rm pz}$. 
           Different lines from top to bottom correspond to different 
           $N_{\rm{pz}}$: $61$ (short dashed line), $31$ (solid line), 
           $21$ (dash 3-dotted line), $11$ (dash dotted line) and
           $6$ (long dashed line).  Note that the results have
           converged with $N_{\rm pz}\ge 21$; we use $N_{\rm pz}=31$ just to be
           conservative.  
  }
% Plotted by cheops3:simplePlotWWaDE.pro
% Plotted by cheops3:simplePlotWWaDEnew.pro
  \label{fig:npz31vs61}
\end{figure}

Figure~\ref{fig:npz31vs61} shows the degradation in the errors on $w_a$ for the
cases of $N_{\rm{pz}}=6$, 11, 21, 31 and 61 as a function of the
prior on the photo-$z$ parameters.   Results for $w_{0}$ are similar.
To compare priors for different $N_{\rm pz}$ values, we have here
rescaled the individual parameter priors by $\sqrt{N_{\rm pz}/31}$ so as to
be equal for a fixed redshift interval $\delta z = 0.1$.
The results have converged with
$N_{\rm pz}\geq 21$.  To be conservative, in the rest of this paper
we use $N_{\rm pz}=31$, or a total of 62 photo-$z$ parameters.

The impact of this choice of $N_{\rm pz}=31$ as a function of $N_{\rm div}$ 
for dark energy parameters is shown in Figure~\ref{fig:numDiv} (points).  
For all 
$N_{\rm div}$,  these constraints match those with no tomographic binning 
very well, showing that without prior information on the photo-$z$ parameters 
all tomographic information has been effectively destroyed and we recover 
the case with a single redshift division.  The small discrepancy comes from 
the inadequacies in the Fisher matrix one of which is the local approximation 
to the parameter errors as we shall discuss in the next section.

\subsection{Photo-$z$ -- Dark Energy Degeneracy}
\nopagebreak
\label{sec:degeneracy}
With a sufficient number of unknown photo-$z$ parameters 
$2N_{\rm pz} \gtrsim 62$, the Fisher matrix results of the previous section 
imply that dark energy information in tomography is completely lost. This 
fact implies that there is a nearly perfect degeneracy between photo-$z$, 
dark energy and other cosmological parameters.  Here we examine that aspect 
of the degeneracy that involves the photo-$z$ and dark energy
parameters only.   This degeneracy alone suffices to destroy most of the
tomographic information and will remain even if the other cosmological 
parameters are perfectly measured from other sources.

Constructed from parameter derivatives, the Fisher matrix is a local expression 
of the degeneracy in parameter space.  Because the Fisher matrix results imply 
that the degeneracy persists to large changes in the dark energy parameters, it 
is important to assess the extent of the degeneracy more directly and test the 
validity of the Fisher approximation.  If the degeneracy relation ``curves" in 
parameter space, the Fisher approximation will only find the local tangent.

We start by identifying this local tangent with the Fisher matrix. To isolate 
the degeneracy between dark energy and photo-$z$ parameters, we eliminate the 
other cosmological parameters, formally by adding strong priors to the Fisher 
matrix.  For numerical reasons we also add a weak prior on photo-$z$ parameters 
($\Delta z_{\rm{bias}} = \Delta \sigma_{{z}} = 1$ ) to control numerical errors 
from the nearly singular Fisher matrix.  Of the eigenvectors of this Fisher 
matrix that involve the dark energy, those with the smallest eigenvalues will 
be responsible for most of the photo-$z$ dark energy degeneracy.
We find that a single linear combination of
parameters (dark energy {\it and} photo-$z$) contributes most ($\sim 98\%$) of
the errors in dark energy parameters $w_0$ and $w_a$. Thus the degeneracy is 
essentially one dimensional in the multi-dimensional parameter space.  
 Let us call this direction -- or the eigenvector of the Fisher matrix 
-- ${\bf e}_w$.

The true extent of the degeneracy is quantified by the change in $\chi^2$ 
between the fiducial model $p_\mu$ and a trial model $\tilde p_\mu$
\ba 
\Delta \chi^2_{\rm true} &=&{} \sum_{\ell=2}^{\ell_{\rm max}}(2\ell+1)f_{\rm sky} \sum_{ab}
    \left [O_{a}(\ell; p_{\mu})-
	O_{a}(\ell; \tilde p_{\mu})\right ]
    \nonumber \\
    &&{} \times  [{\bf C}^{-1}]_{ab}
	      \left [O_{b}(\ell; p_{\mu})-
		  O_{b}(\ell; \tilde p_{\mu})\right] \,.
\ea
If the Fisher matrix approximation were valid out to say $1\sigma$ along the
degeneracy then the trial model $\tilde p_\mu = p_\mu + \sigma_w {\bf e}_w$, 
where $\sigma_w^{-2}$ is the eigenvalue corresponding to ${\bf e}_w$, would be
separated by  
\ba 
\Delta \chi^2_{\rm F} 
    &= & \sigma_w^2 {\bf e}_w^T\, {\bf F}\, {\bf e}_w=1\label{EQ:chi2}
\ea 
due to the orthonormality of the eigenvectors. In practice 
$\Delta \chi^2_{\rm true}=857$ for this extrapolation indicating a curvature
in the degeneracy direction.    In other words, $p_\mu$ and $\tilde p_\mu$ are
highly distinguishable models in spite of the Fisher prediction that they are 
indistinguishable.

Even given curvature in the degeneracy direction, the Fisher approximation 
remains useful if it accurately predicts the extent of the degeneracy.  This 
is especially true if the curvature lies mainly in the photo-$z$ nuisance 
parameters which exist only to be marginalized.    To assess the extent of 
the degeneracy, we use  the Fisher matrix as a local approximation of the 
degeneracy with the following procedure.  Starting at the fiducial model, 
calculate the Fisher matrix and find ${\bf e}_w$ as defined above, then take 
a {\it small} step along ${\bf e}_w$ direction. Now calculate the Fisher 
matrix at the new point, find the new ${\bf e}_w$,  take another small step 
along this new direction.  Repeat.  
The smallness of the steps is controlled such that the change of $\chi^2$ 
between steps agrees within $10\%$ with that predicted by Fisher matrix. 
This $10\%$ error may add up to a much bigger percentage after a few steps.
To make sure that we stay on the degeneracy direction, local minimum of 
$\Delta \chi^2_{\rm true}$ is searched after a few steps out.

\begin{figure}[ht]
\centerline{\psfig{file=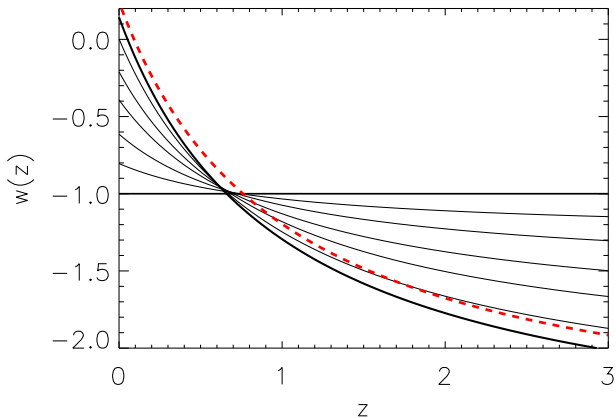, width=3.2in}}
  \caption{Equations of state of dark energy $w(z)$ that are degenerate with
     photo-$z$ parameters.  Solid lines show a series of degenerate models 
     stepping out by $\Delta w_0=0.2$ and ending at a model with 
     $\Delta w_0=1.14$ which deviates from the fiducial model at $+1\sigma$ for 
     the fiducial survey.   The tight correlation between $w_0$ and $w_a$ 
     along the degeneracy direction results in the tight ``waist'' or pivot
     where $w(z)$ remains well determined.  The dashed line shows that the
     $+1\sigma$ degenerate model as predicted by the Fisher matrix is in 
     good agreement with the true degeneracy even out to large $\Delta w_0$.  }  
  % Plotted by rhoDEnew.pro(cheops3)
  \label{fig:wz}
\end{figure}

From this construction we find that the extent of the degeneracy in $w$ is 
accurately predicted by the Fisher matrix. Figure~\ref{fig:wz} 
shows that the model with $\Delta \chi^2_{\rm F}=1$ (thick dashed line)
is almost identical to the 
model with $\Delta \chi^2_{\rm true}=1$ (thick solid line) in $w(z)$.
In Figure~\ref{fig:wz} we also show intermediate models along
the degeneracy with $\Delta \chi^2_{\rm true}<1$.  That they all pass through 
essentially a single point in $w(z)$ space is another indication that the
 degenerate direction lies almost entirely along a specific linear 
combination of dark energy parameters as predicted by the Fisher matrix.
The curvature in parameter space mainly involves the photo-$z$ parameters. 

The redshift at which these curves intersect is $z\approx 0.7$ and at this
redshift measurements of $w$ are essentially immune to photometric
redshift errors.  This immunity reflects the fact that even without 
tomography lensing can constrain the equation of state at some effective 
redshift.  With two parameters to describe $w(z)$, there is only one 
remaining linear combination to be affected by photometric redshifts.
With a more general parameterization of $w(z)$ we expect that there will be
multiple degenerate directions with roughly the same single aspect of $w(z)$ 
preserved.

\begin{figure}[ht]
  \centerline{\psfig{file=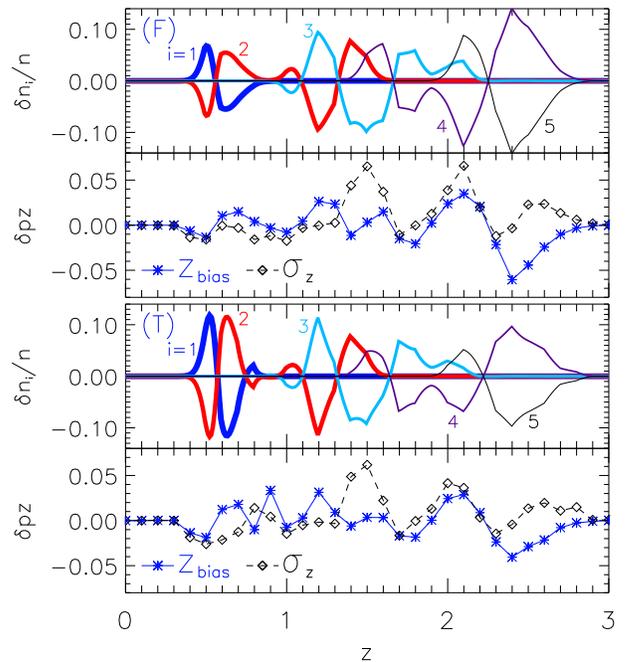, width=3.2in}}
  \caption{Comparison of the fractional differences of the redshift 
           distribution $\delta n_i/n$ in each of the $i=1,\ldots,5$ tomographic bins that establishes 
           the 1$\sigma$ degeneracy with dark energy parameters of an extent 
           $\Delta w_{0}\sim 1$.  Also shown are the changes  
           in the two sets of discrete photo-$z$ parameters $\delta$pz = $(\delta \sigma_z$, 
           $\delta z_{\rm bias})$ that cause these differences,
           connected by lines to guide the eye.
           Top panel (F): Fisher approximation for the $1\sigma$ degeneracy.
           Bottom panel (T): true degeneracy.
           }
         % Plotted by nzDegenerate.pro(cheops3)
  \label{fig:nz2}
\end{figure}

Furthermore, the two estimates (true and Fisher) agree on the  amplitude
of the photo-$z$ parameter variation along the degenerate direction. For 
example, as shown in Figure\,\ref{fig:nz2}, at a 
point along the ${\bf e}_w$ direction which is 1$\sigma$ away from the fiducial
model, the Fisher matrix indicates that the photo-$z$ parameters changed by 
$\delta z_{\rm{bias}} < 0.06$ and $\delta \sigma_{{z}} < 0.07$, while the 
actual bounds on the variations are $\delta z_{\rm{bias}} < 0.04$ and $\delta
\sigma_{{z}} < 0.07$.    Note that these changes are fairly small and imply 
that subtle variations in the redshift distributions for the tomographic bins
are responsible for a degeneracy that degrades errors in $w_0$ and $w_a$ by
an order of magnitude.  
Figure~\ref{fig:nz2} also shows that the difference between these distributions
for $\Delta \chi^2_{\rm true}=1$. We expect that with a change in the photo-$z$
model, the specific photo-$z$ variations that establish this degeneracy 
will change but that a strong degeneracy will remain.

In summary, we find that the Fisher matrix is an adequate tool for assessing
the existence and extent of degeneracies between photo-$z$ and dark energy
parameters.   It should not however be used to infer the specific changes in
the photo-$z$ parameters that establish the degeneracy.   

\section{Photo-$z$ Information Recovery}
\label{sec:results}

In the previous section, we established the existence of a degeneracy 
between photo-$z$ parameters and dark energy parameters and tested the 
validity of the local Fisher matrix approximation to this degeneracy.   
In this section, we use the Fisher matrix formalism to investigate
the extent to which prior information on the photo-$z$ distributions 
help recover the tomographic dark energy information.   
We assume $2N_{\rm pz}=62$ photo-$z$ parameters and $N_{\rm div}=5$ 
tomographic bins through out this section (see \S \ref{sec:information}).

\begin{figure}[ht]
\centerline{\psfig{file=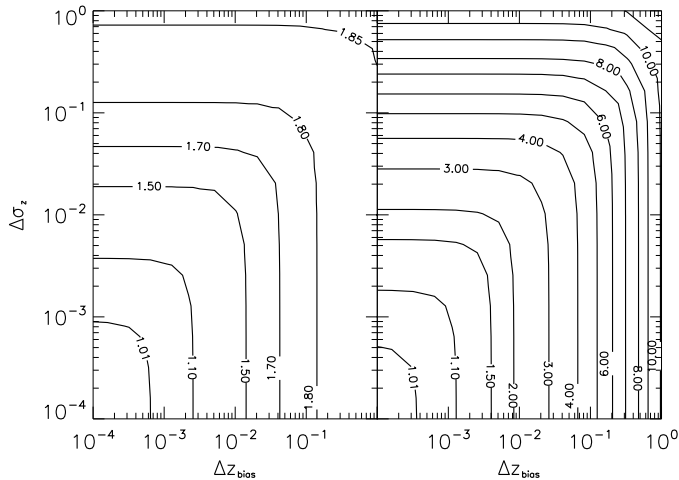, width=3.5in}}
  \caption{Error degradations for constant $w_0$ (left panel) and those for $w_a$
           when both $w_0$ and $w_a$ are varied (right panel), as a
           function of photo-$z$ parameter priors. Here the degradations are
           defined as actual errors relative to errors that assume the photo-$z$
           parameters to be perfectly known. Priors on the photo-$z$ parameters
           $z_{\rm{bias}}(z_{\mu})$ and $\sigma_{{z}}(z_{\mu})$ are shown on the
           x-axis and y-axis respectively. Fiducial photo-$z$ model I is assumed
           and the photo-$z$ parameter spacing in redshift is $\delta z = 0.1$.
            } 
% ploted by finalPZpriorCon.pro(cheops3)
  \label{fig:main}
\end{figure}

\subsection{Photo-$z$ Priors}
\label{sec:priors}

We now explore the effect  on dark energy parameter constraints of priors 
on each of the photo-$z$ parameters $z_{\rm bias}(z_{\mu})$ 
and $\sigma_z(z_{\mu})$.  For simplicity,
we begin by applying a redshift independent prior on the parameters.
In practice parameters controlling the distributions well above and
well below the median redshift require weaker priors.  We will discuss
this point in \S \ref{sec:training}.

In Figure~\ref{fig:main}, the left panel shows the error degradation in 
$w_0$ assuming the $\{w_0, \Omega_{\rm DE}\}$ parametrization (left panel) and 
$w_a$ assuming the $\{w_0, w_{a}, \Omega_{\rm DE}\}$ parametrization 
(right panel). For reference the baseline errors for the fiducial survey  
are listed in Table\,\ref{tbl-1}. 

As in the previous section where other cosmological parameters were 
artificially fixed, we find that the larger dark energy parameter space 
is more susceptible to photo-$z$ errors. For example, for the extreme 
case of no photo-$z$ information, i.e.\ very weak priors on both bias 
and scatter parameters, dark energy parameters of the 
$\{w_0, \Omega_{\rm DE}\}$ parameterization are
degraded by about a factor of two while those of 
$\{w_0, w_{a}, \Omega_{\rm DE}\}$ parameterization are degraded by 
about a factor of ten.

In the more relevant case where we demand that the dark energy error 
degradation be no larger than $1.5$, the requirement per photo-$z$ 
parameter is about $0.01$ for $\{w_0, \Omega_{\rm DE}\}$ and $0.003$ 
for the $\{w_0, w_{a}, \Omega_{\rm DE}\}$ parametrization.  
Figure~\ref{fig:main} also shows that both bias and scatter parameters
are important, and that knowledge of the bias is only slightly more important
than that of the scatter. Furthermore, dark energy parameters that are not
shown in Figure~\ref{fig:main} have very similar requirements to those plotted
in either parametrization.

\subsection{Dependence on Fiducial Model}
\label{sec:variations}

\begin{figure}[ht]
\centerline{\psfig{file=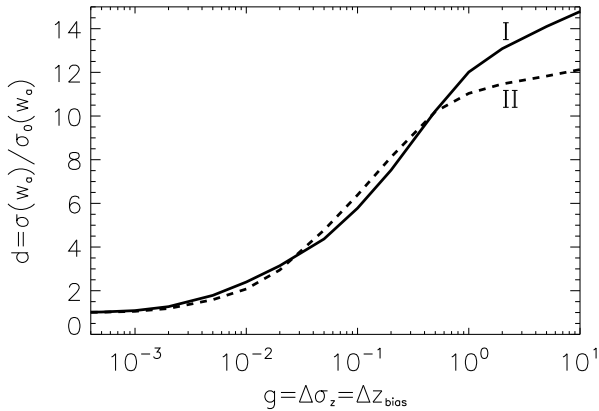, width=3.2in}}
  \caption{Comparison of the photo-$z$ requirements for our two fiducial 
           photo-$z$ models.  The solid line shows the degradations for 
           model I while the dashed line corresponds to model II.  The 
           fiducial errors in $w_a$ when the photo-$z$'s are perfectly 
           known are $\sigma_0(w_{a}) = 0.69$ (model I) and 
           $\sigma_0(w_{a}) = 0.96$ (model II). }
% cheops4: 0112 and 0116 simplePlot.pro
% cheops3: 0407 and cheops1_0407 simplePlot0.pro
  \label{fig:PZIvsII}
\end{figure}

The results above are for a specific choice of the fiducial model for
the photo-$z$ distribution and survey. To explore the dependence on the former
we take the very different photo-$z$ Model II, where the scatter 
is substantially larger and jumps discontinuously in redshift.
Even so the requirements on the photo-$z$ parameters are very similar; see 
Figure~\ref{fig:PZIvsII}. In particular, within the interesting regime where 
the photo-$z$ prior is smaller than unity and the degradation in dark energy 
parameter errors is a factor of a few or lower, the two models agree very well.

On the other hand, photo-$z$ requirements do depend on the parameters
$f_{\rm sky}$, $\gamma_{\rm int}$ and ${n}^A$
that determine the level of sample and noise variance in the survey.
The trend is that the more ambitious the survey, the more stringent the 
requirements on photo-$z$ parameters. The scaling for the
priors on photo-$z$ parameters  can roughly be described as
\begin{eqnarray}
\label{eq:scale}
 {\Delta p (f_{\rm sky},\gamma_{\rm int}, n^{A}, d)  \over 
   {\Delta p (0.1,0.4,55,1.5)}}
   &=& {g(d)\over g(1.5)} \sqrt{ 0.1 \over f_{\rm sky} }  \\
    && \times
      \left[   1 + C \left(
       {\gamma_{\rm int}^{2}\over 0.16 } 
       {55 \over {{n}^A} } -1  \right) \right]\,, \nonumber 
\end{eqnarray}
where $\Delta p=\Delta z_{\rm bias}= \Delta \sigma_{z}$ 
gives the photo-$z$ priors and $n^{A}$ is in units of arcmin$^{-2}$.
Here $g(d)=\Delta p(0.1,0.4,55,d)$ scales the prior requirement to alternate
levels of degradation $d$; it is shown in Figure~\ref{fig:PZIvsII} as 
$d(g)$  and is only weakly dependent on the fiducial
photo-$z$ model. With the best fit $C = 0.6$ this scaling 
is good up to a factor of 2 for any reasonable set of survey parameters.  

Finally the photo-$z$ precision requirement is not very sensitive
to $z_{\rm med}$, the median source redshift of the survey. For 
$0.68 < z_{\rm med} < 1.3$, $\Delta p$ varies by less than $40\%$.

\subsection{Training Set Size} \label{sec:training}

The priors on the photo-$z$ parameters ultimately require
a training set of galaxies with measured spectroscopic redshifts.
Operationally suppose that a photo-$z$
training set has $N_{\rm spec}^{\mu}$ spectroscopic redshifts per 
redshift interval   determined by $N_{\rm pz}$ (here $\delta z=0.1$).  

Given a Gaussian distribution for the photo-$z$ distribution and a fair 
sample of spectroscopic galaxies selected from this distribution, the 
training set would independently determine the bias and scatter to 
\begin{eqnarray}
\Delta z_{\rm bias}(z_{\mu})&=&\sigma_z(z_{\mu})\sqrt{1 / N_{\rm spec}^{\mu}} \,\, ,\nonumber\\
\Delta\sigma_z(z_{\mu}) &=& \sigma_z(z_{\mu}) \sqrt{2 /N_{\rm spec}^{\mu}}\,\, .
\label{eq:nspect}
\end{eqnarray}
For a fixed dark energy degradation, $N_{\rm spec}^{\mu}$ depends on 
two things: the fiducial $\sigma_z$ and the required prior as scaled from
 equation~(\ref{eq:scale}).
Since the photo-$z$ prior requirement is roughly independent of $\sigma_z$
as shown in Figure\,\ref{fig:PZIvsII}, the larger the scatter, the larger the
required training set.  
Note that $N_{\rm spec}$ ($ \equiv \sum_{\mu} N_{\rm spec}^{\mu}$ ) is 
robust to changes in the number of photo-$z$ parameters or $\delta z$.
For example a binning of $\delta z =0.05$ would imply twice
as many photo-$z$ parameters that would need to be constrained a factor of
$\sqrt{2}$ less well, yielding the same requirement on $N_{\rm spec}$.

For determining the redshift extent of the training set, it is important to
go beyond our simple redshift independent prior.  Figure\,\ref{fig:nspect} 
shows the cumulative $N_{\rm spec}(>z)$ required for 1.5 degradation in 
dark energy. Notice that this flat prior assumption would require a 
substantial number of galaxies across the whole redshift range ($10^5$), 
including $8 \times 10^4$ galaxies above $z=1.5$. This number is 
artificially high since the actual requirements on the prior fall sharply 
away from the median redshift of the distribution. 

\begin{figure}[ht]
\centerline{\psfig{file=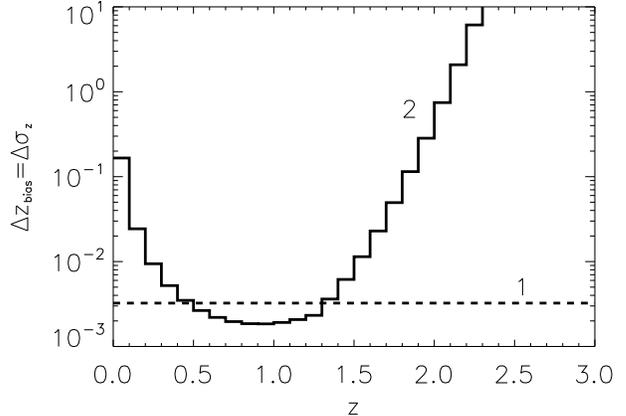, width=3.2in}}
  \caption{Photo-$z$ prior templates of how the bias and scatter priors vary 
           with redshift to produce a fixed  error degradation in dark energy 
           parameters. The degradation on $w_{a}$ is $1.5$.
           The straight line (1; dashed) is the flat template used in 
           Figure\,\ref{fig:main}. The curved line (2; solid) is the weighted 
           template constructed according to the galaxy number density. 
           At $z<1.2$, the curve follows ${\bar n}_i^{-1}$
           while at $z>1.2$, the curve follows ${\bar n}_i^{-3}$. Here
           ${\bar n}_i$ is the number of galaxies in each redshift interval
           $\delta z = 0.1$.  }
  % Plotted by zbiasHisto2.pro(cheops3)
  \label{fig:zbiashisto}
\end{figure}

\begin{figure}[ht]
\centerline{\psfig{file=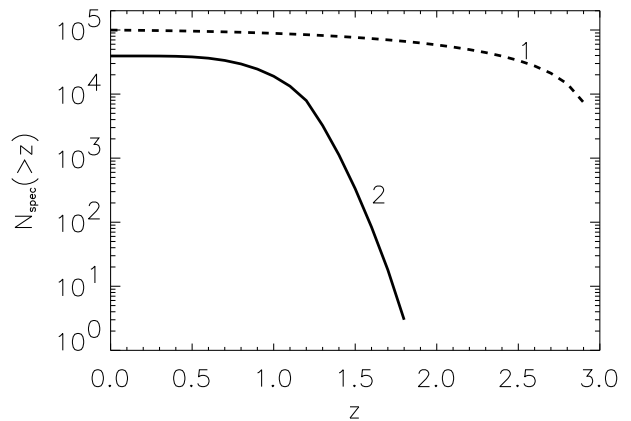, width=3.2in}}
  \caption{Cumulative requirement of $N_{\rm spec}$ for $1.5$ dark energy 
          error degradation. The corresponding photo-$z$ prior templates are 
          shown in Figure\,\ref{fig:zbiashisto} with (1; dashed) as the flat 
          prior and (2; solid) as the weighted prior.}
  % Plotted by Nspect.pro(cheops3)
  \label{fig:nspect}
\end{figure}

To illustrate the difference, we constructed a weighted template of how 
the bias and scatter priors vary with redshift to produce a fixed 
degradation in dark energy parameters.  We choose a  simple power 
law of ${\bar n}_i$ which is the number of galaxies in each redshift 
interval ($\delta z = 0.1$ for the fiducial choice of $N_{\rm pz}=31$) 
corresponding to the photo-$z$ parameters. At $z < 1.2$, the power is 
chosen as $-1$.  To account
for the difficulty in measuring redshifts at $z > 1.5$ from optical bands, 
we steepen the index to  $-3$ for $z>1.2$.
Figure\,\ref{fig:zbiashisto} compares the flat prior to the weighted one.
The requirement for the weighted one drops to a total of 
$4 \times 10^4$  but more importantly only  $300$ at $z>1.5$. 

\begin{figure}[ht]
\centerline{\psfig{file=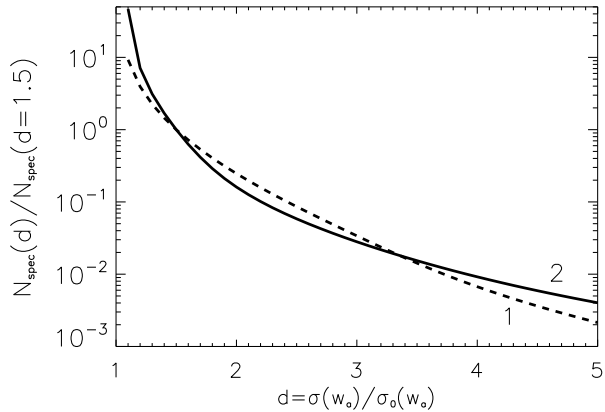, width=3.2in}}
  \caption{The ratio of required $N_{\rm spec}$ for an arbitrary $w_a$ 
           degradation relative to that of $1.5$ degradation for the two 
           prior templates of Figure\,\ref{fig:zbiashisto}. } 
         % as in Fig.\,\ref{fig:nspect}.  }
  % Plotted by fit.pro(cheops3)
  \label{fig:nspectRatio}
\end{figure}

For dark
energy degradations other than 1.5, we provide in Figure\,\ref{fig:nspectRatio}
the ratio of $N_{\rm spec}$ for an arbitrary dark energy degradation to that 
of 1.5 degradation. In order to find out the $N_{\rm spec}$ requirement for 
any dark energy degradation, all one needs to do is to look up the ratio in 
Figure\,\ref{fig:nspectRatio} and multiply it by the $N_{\rm spec}$ in
Figure\,\ref{fig:nspect}.

We have tested that the scaling relation to surveys with different fiducial 
parameters of  equation~(\ref{eq:scale}) works equally well for both the flat 
and weighted priors. Using the scaling relation, the requirement of 
$N_{\rm spec}$ could be scaled to different survey easily.

\subsection{Mean vs. Median}
\label{sec:mean}

We now estimate the amount of information that comes from knowing the 
median or mean of the redshift distribution of source galaxies in each 
tomographic bin.  Note that this is distinct from priors on the 
photo-$z$ bias or mean photo-$z$ at a given redshift.  
This question is interesting in its
own right, but also because other work, parallel to this \citep{wlsys}
has parametrized the photo-$z$ uncertainty by the centroids
of the tomographic bins, which when varied shift the overall distribution of
the corresponding tomographic bin.  While one intuitively expects that the
centroid  of the photo-$z$ bin (or, more generally its mean)
carries the most information, we now have the tools
to precisely examine the relative contribution of the mean relative to that 
of the higher moments.

\begin{figure}
%\centerline{\psfig{file=../ps/mean.ps, width=3.2in,height=2.4in}}
\centerline{\psfig{file=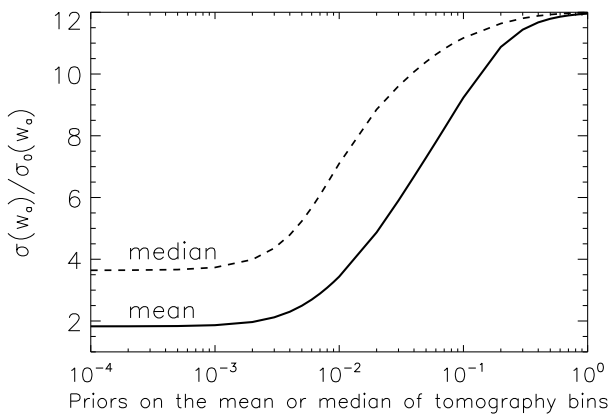, width=3.2in}}
  \caption{The effect of the mean or median of tomography bins. Vertical
           axis is dark energy error degradation. The solid line and 
           the dashed line are for the case of mean and median respectively. 
           Photo-$z$ priors of $\Delta \sigma_z = \Delta z_{\rm bias} = 1$
           are assumed. }
    %plotted by finalPZpriorCon2.pro(cheops3).
    %plotted by M1vsMed.pro(cheops3).
  \label{fig:mean}
\end{figure}

Figure~\ref{fig:mean} shows degradations of the error in $w_a$ as a function of 
priors on the mean or median of the tomographic bins. All of the photo-$z$ 
parameters are given a weak prior of unity for numerical stability. A prior 
of $10^{-3}$ on the mean is enough to render the mean essentially precisely 
known. But the dark energy degradation is still a factor of  2 even with 
perfect mean measurements showing that there remains information lost to the 
higher moments of the distribution.

Similarly, while the mean of the tomographic distribution does carry the 
majority of the information, the median carries significantly less
(see Figure\,\ref{fig:mean}).  The reason is that the mean has extra 
information about the tails of the redshift space distribution while
the median does not. This sensitivity to the tails will make obtaining precise
measurements of the mean difficult.   One still requires a fair sample 
from each of the tomographic redshift bins extending to high redshift. 
In the end, mean priors require 
a similar  number of training set galaxies as above.

\section{Discussion}

\label{sec:discussion}

We have performed a systematic study of the effects of imperfect photometric
redshifts on weak lensing tomography. Describing the photo-$z$ distribution 
with a bias and scatter that can vary arbitrarily between redshift intervals of 
$\delta z = 0.1$, we 
studied the degeneracies between photo-$z$ and dark energy parameters, as 
well as the resulting degradations in dark energy parameter errors.

Not surprisingly, we find that there exist significant degeneracies
between the dark energy and photo-$z$ parameters. Assuming that the overall
distribution of galaxies $n(z)$ is independently known and the photometric
redshifts are used only for tomographic subdivision, we find that larger dark
energy spaces suffer more degeneracy with photo-$z$ than the smaller ones.  

Without any information on
photo-$z$ parameters, one recovers the no tomography case where errors
on fiducial parameters are a factor of two times worse (for the $\{w_0,
\Omega_{\rm DE}\}$ parametrization) or ten times worse 
(for the $\{w_0, w_{a}, \Omega_{\rm DE}\}$ parametrization) than 
those for the 10-bin tomography case with perfect photo-$z$'s.

For the fiducial survey, in order to have less than a factor of $1.5$ 
degradation in dark energy parameter errors, the uncertainties of 
photo-$z$ parameters $z_{\rm bias}$ and $\sigma_{z}$ (defined in the
redshift interval $\delta z = 0.1$) should each be controlled to better than
$0.003$-$0.01$, depending again on the size of dark energy parameter space. 
We provide a convenient approximation for scaling these requirements to 
different surveys.  Importantly, no single number such as the mean or median 
of galaxies in the tomographic bin captures all of the effect of photo-$z$ 
errors.  That the mean captures more of the information than the median 
indicates that the dark energy information is sensitive to the tails of the 
distribution.

In order to achieve less than a factor of $1.5$ degradation in the evolution 
of the equation of state, a training set of a few times $10^4$ galaxies with 
spectroscopic redshifts is required. Again, one can easily rescale the number 
of galaxy requirement to different surveys using our scaling relation.

There are several caveats to our assessment that merit future study.
Although our parametrization can handle photo-$z$ degeneracies, for example
from multiple galaxy types,
that lead to bimodality and catastrophic errors, we have limited our study to 
fiducial models around which their  effects are small.  
Such effects will increase the required number of training set galaxies.
Moreover, we have assumed that the parent redshift distribution of the survey 
is known and that photometric redshifts are only employed to subdivide
the galaxy sample for tomography.  Uncertainties in the parent distribution can
further degrade dark energy determinations.   

On the other hand, uncertainties in the parent distribution are also 
constrained by the training set.  If one assumes that $n(z)$ is a smooth 
function that is parametrized by relatively few parameters, uncertainties
in the parent distribution should be smaller than those of the tomographic bins.
For illustrative purposes, if we parametrize $n(z)$ with the
three parameters of  equation~(\ref{eq:nz}),
we find that the constraint on $n(z)$ from $N^{\mu}_{\rm spect}$ is good enough
to have dark energy parameter errors differ by less than $10\%$ from the
case where $n(z)$ is fixed.

Given the current state-of-the-art of photo-$z$ algorithms as well as expected
improvements with multi-wavelength observations of all source galaxies,
prospects for sufficiently accurate determination of photometric redshifts are
bright. Nevertheless, it will be an important and challenging task to achieve
good control of the photo-$z$ accuracy for the specific types of galaxies
selected in lensing surveys, and then  propagate the
remaining photo-$z$ errors into the final cosmological constraints.

\acknowledgements {\it Acknowledgments}: We thank Carlos Cunha, Eric Gawiser, 
David Johnston, Marcos Lima, Takemi Okamoto, Hiroaki Oyaizu, Eduardo Rozo and 
Tony Tyson for useful discussions. ZM and WH are supported by the Packard 
Foundation and the DOE. DH is supported by the NSF Astronomy and Astrophysics 
Postdoctoral Fellowship under Grant No.\ 0401066.

%\include{DBbib}

%\end{doublespace}

\end{document}